\documentclass[aps,prc,onecolumn,superscriptaddress,
showpacs,showkeys]{revtex4}
\usepackage{graphicx}
\usepackage{float} 
\usepackage{bm}
\usepackage{epsfig}
\usepackage{amsfonts}
\usepackage{amsmath}
\usepackage{amssymb,amscd}
\usepackage{xcolor}
\usepackage{ulem}

\begin{document}
	
\title{A study of charged-particle multiplicity distribution in high energy p-O collisions}

\author{Yuri N. {\sc Lima}}
\email{ylima@if.usp.br}
\affiliation{Instituto de F\'{\i}sica - Universidade de S\~ao Paulo
Rua do Mat\~ao 1371 - CEP 05508-090\\
Cidade Universit\'aria, S\~ao Paulo - Brasil}

\author{Lucas J. F. {\sc Silva}}
\email{lucasj.silva@usp.br}
\affiliation{Instituto de F\'{\i}sica - Universidade de S\~ao Paulo
Rua do Mat\~ao 1371 - CEP 05508-090\\
Cidade Universit\'aria, S\~ao Paulo - Brasil}

\author{André V. {\sc Giannini}}
\email{andregiannini@ufgd.edu.br}
\affiliation{
  Faculdade de Ciências Exatas e Tecnologia, Universidade Federal da Grande Dourados, 
  Dourados, MS, Brazil, 79804-970
}
\affiliation{Departamento de F\'isica, Universidade do Estado de Santa Catarina, 89219-710 Joinville, SC, Brazil}

\author{Marcelo G. {\sc Munhoz}}
\email{munhoz@if.usp.br}
\affiliation{Instituto de F\'{\i}sica - Universidade de S\~ao Paulo
Rua do Mat\~ao 1371 - CEP 05508-090\\
Cidade Universit\'aria, S\~ao Paulo - Brasil}
	
\begin{abstract}

This study investigates the multiplicity distribution of charged particles generated in $p$-O collisions, employing Pythia (Angantyr) and $k_T$-factorization approach. Oxygen nucleus configurations are sampled using a $\alpha$-cluster model to evaluate both formalisms and assess how initial nucleus configuration influences the properties of the produced final states. Results obtained through clustering are systematically compared to those derived from the Woods-Saxon nuclear distribution. The analysis encompasses various pseudorapidity intervals ($|\eta|<$ 0.5, 1.0, 2.0, 3.0) and center-of-mass energies ($\sqrt{s}=$ 2.36, 5.02, 7.0, 13.0 TeV). Based on the resulting distributions, we examine the KNO scaling effect and fit the distributions with the double NBD model for parameterization, aiming to accurately characterize the observed results and elucidate contributions from both soft and semi-hard processes. Our results indicate that different geometric descriptions of the oxygen nucleus project significantly different multiplicities of charged particles, especially for large multiplicities and higher pseudorapidity. We also observed that multiplicity of charged particles calculated with Pythia reveals significantly different behavior from that calculated with $k_T$-factorization.

\end{abstract}

\date{\today}
\maketitle

\section{Introduction}

There is an increasing interest in the investigation of oxygen systems, particularly following the recent introduction of oxygen beams into Run 3 of the Large Hadron Collider (LHC). This development presents new opportunities to examine particle production from smaller systems within a previously unstudied energy domain. Traditionally, lead ions have been utilized at the LHC for heavy ion collisions, as they are believed to create the optimal conditions for Quark-Gluon Plasma (QGP) formation. In contrast, oxygen nuclei are significantly smaller than lead nuclei, offering valuable comparative insights when studying nuclear collisions in relation to the size of the colliding nuclei. Additionally, data from O-O and $p$-O collisions at high energies could enhance our understanding of collective phenomena emergence and partonic energy loss in small systems. These findings may also inform models describing interactions between high-energy cosmic ray protons and atmospheric nuclei, thereby improving the precision of atmospheric shower simulations that rely on hadronic interaction models at energy scales beyond those accessible with fixed-target experiments.

Recent research indicates that the geometric configuration of the oxygen nucleus -- whether represented by a continuous Woods-Saxon distribution or as a compact array of tetrahedral $\alpha$-clusters -- plays a significant role in final-state particle production \cite{Ding2023,Sharma2025}. Employing models that isolate purely geometric effects from initial conditions can enhance our understanding of how the structural organization of the nucleus influences particle yields. Thus, comparing a clustered structure with a smooth Woods-Saxon distribution suggests that the chosen description of the nucleus substantially affects outcomes in particle production.

The literature contains extensive data on multiplicity distribution, particularly concerning charged particles \cite{ALICE2017, ALICE2025, CMS2011, CMS2018, ATLAS2011, ATLAS2016}, alongside various phenomenological studies addressing this subject \cite{Duan2025, Islam2025, Dokshitzer2025JHEP10, Dokshitzer2025JHEP08, Kulchitsky2023, Levin2024, Grosse2009, MartinsFontes2025PRD112, MartinsFontes2025Physics7, Germano2021, Germano2024}. Particle multiplicity serves as a global observable that facilitates event characterization in colliding systems and yields valuable insights into the dynamics of hadronic interactions at high energies. As such, this observable supports the investigation of qualitative shifts in reaction dynamics. To date, existing experimental data have primarily led phenomenological research on multiplicity to focus on $pp$ or $p$-Pb collisions \cite{ALICE2017, ALICE2025}. However, with recent advancements in $p$-O collision data, this study provides a comprehensive analysis of the multiplicity distribution for related particles in these collisions, examining their fundamental properties and relevant aspects.

This work is structured as follows: in Section \ref{sec2} we introduce a summary of the formalisms used in this study and list the main parameters used; Section \ref{sec3} provides details of the nucleus description using both the $\alpha$-cluster and the Woods-Saxon; in Section \ref{sec4} we discuss the charged-particle multiplicity distribution in $p$-O collisions and present the results and observations regarding this distributions; in Section \ref{sec5} we implement the KNO scaling in the obtained distributions; in Section \ref{sec6} we present the parameterization of the distributions with the double NBD; and we conclude by summarizing the main results obtained.

\section{Event simulation and approaches}
\label{sec2}

Pythia, initially designed for $pp$ systems, is one of the most widely algorithm used today and simulates particle production from collisions with focus on high energies. The framework seeks to reproduce the particle production process from the interaction of the initial states to the final particles and record the information 
about their momenta, energies, and generation history. The simulation follows a sequence of steps that seek to model the main known physical phenomena of particle production: 
$pp$ collisions involve one or more perturbative scattering processes between the incident partons, which are implemented within the MultiParton Interactions (MPI) framework \cite{Sjostrand1987}; the perturbative scattering processes are accompanied by Initial-State Radiation (ISR) and Final-State Radiation (FSR); the next step is the Lund string fragmentation model \cite{Andersson1983} to implement hadronization, where the created partons and the beam remnants are connected by strings (color flux tubes). When the partons move away the string breaks, generating a light quark-antiquark pair. This repeats until string fragments large enough to be identified as hadrons remain. However, when we consider Color Reconnection \cite{Argyropoulos2014}, the strings are reorganized, reducing their total length, allowing partons to connect with each other. 
A extensive discussion and further information regarding underlying processes of the simulated interaction can be found in the Reference \cite{Pythia8.3}.

Pythia adopts the Angantyr model \cite{Bierlich2018} to handle collisions with nuclei, thus enabling $p$-A and A-A simulations but without considering the formation of QGP matter. Angantyr model considers the Woods-Saxon nuclear charge density distribution to randomly manifest the positions of nucleons within the interacting nuclei, and the Glauber formalism to obtain the number of nucleon-nucleon interactions. The latter is based on the eikonal approximation in the impact parameter space, assuming that projectile nucleons move in straight lines and subsequently undergo multiple subcollisions with target nucleons. Furthermore, corrections for diffractive excitation, resulting from fluctuations in the nuclear substructure, and fluctuations in the cross-section of the target and projectile nucleons are accounted via Glauber-Gribov color fluctuation model. It is still necessary to consider the contribution to the final state of each nucleon-nucleon interaction; this is done using the Fritiof model, which is based on nucleons hit both non-diffractively and diffractively (contributing to soft production). Events with multiple non-diffractive nucleon-nucleon subcollisions are further characterized into primary and secondary non-diffractive interactions based on impact parameters. 

The description above indicates that Pythia approximates and has a conceptual affinity with the structure given in the so-called ``collinear factorization'', which allows deriving cross sections in terms of integrated parton distributions, solutions of linear evolution equations such as the DGLAP equation, and can be succinctly and schematically expressed for the $2\to 2$ process (Pythia considers others) by 
\begin{equation}
\frac{d\sigma^{AB\to hX}}{dyd^2p_T} =\ K\sum_{abcd}\int\frac{dz}{z^2}\int\ dx_adx_b\ f_a(x_a,Q^2)\ f_b(x_b,Q^2)\ \frac{1}{\pi z}\frac{d\sigma^{ab\to cd}}{d\hat{t}}\ D_{i/h}(z,\mu^2),
\label{Eq_fatcoll}
\end{equation}
where this expression considers the convolution of the Parton Distribution Functions $f_{a,b}(x_{a,b},Q^2)$ with the cross-sections of the partonic subprocesses $\hat{\sigma}(ab\to cd)$ and the Fragmentation Function $D_{i/h}(z,\mu^2)$. Furthermore, $x_{a,b}=(p_T/\sqrt{s})e^{\pm y}$ are the momentum fractions that the partons carry, $Q^2$ the momentum transferred during the hard interaction, $z$ is the fraction of the momentum of the light cone of the parton $i$ carried by the hadron $h$, $\mu^2$ is the factorization scale, and $K$ carries the corrections of the Next-to-Leading Order. Some fundamental differences between Pythia and the standard collinear factorization scheme (Equation \ref{Eq_fatcoll}) should be highlighted: in Pythia, hadronization is performed using the non-perturbative Lund model \cite{Andersson1983,Andersson1983jt} and not through an analytical factorization of the type $D_{i/h}$; Pythia relies on collinear factorization for calculating the perturbative part (hard process and showers), but extrapolates by introducing the modeling of the initial transverse moment of the partons, which translates into an effective term, the primordial $k_T$, for the non-perturbative effects (and even for possibly perturbative). The primordial $k_T$ is then an effective term calculated via a two-dimensional Gaussian distribution (details can be found in Reference \cite{Pythia8.3}). This phenomenological approach gives Pythia strong predictive power. 

While the particle production can be estimated via collinear approaches, this description disregards non-linear effects in QCD dynamics that may emerge at high-energies (or, equivalently, at small-$x$) due to the steep rise of partons distributions in the hadronic wave-function, as shown by lepton-hadron collision data~\cite{Cooper-Sarkar:2012vns,Perez:2012um}. Over the years the theoretical description of hadrons in the small-$x$ regime has been consolidated into the Color Glass Condensate effective field theory~\cite{Jalilian-Marian:1996mkd,Jalilian-Marian:1997jhx,Jalilian-Marian:1997ubg,Jalilian-Marian:1998tzv,Kovner:2000pt,Weigert:2000gi,Iancu:2000hn,Ferreiro:2001qy}, which accounts for the parton saturation phenomena in order to restore the unitarity of the scattering process. The onset of the non-linear effects is characterized by a dynamically generated momentum scale~\cite{McLerran:1993ni,McLerran:1993ka}, the saturation scale, $Q_s$, identified as the average transverse momentum in the hadronic wave-function.

The dominant contribution for particle production around midrapidity in high-energy hadronic scatterings is the gluon-gluon interaction and calculations in the CGC framework are based on the $k_T$-factorization formalism, where the invariant cross-section for inclusive gluon production is given by~\cite{Gribov1983,Kovchegov2002,Braun:2000bh},
\begin{equation}
	\frac{d\sigma^{AB\to hX}}{dyd^2p_T}\ = \frac{2}{C_F} \frac{K}{k_T^2} \int\frac{dz}{z^2}\int d^2q_T\,\alpha_s\,\phi_{A}(q_T,y)\phi_{B}(|k_T-q_T|,Y-y)\ D_{g/h}(z,\mu^2)\,,
	\label{Eq_fatkT}
\end{equation}
where $y$ and $k_T$ are, respectively, the rapidity and transverse momentum of the produced gluon and $Y$ is the total rapidity range of the collision; $q_T$ and $|k_T-q_T|$ are the momentum scales of the interacting gluons in the scattering process. The normalization carries a $K$-factor that takes into account higher-order corrections and will be fitted to data. The non-linear, dynamical effects of this interaction are encoded in the Unintegrated Gluon Distributions (UGDs), that correspond to the number of gluons per unit transverse area and per unit transverse momentum in the hadron, for the projectile and the target, $\phi_{A,B}$, which are defined as 
\begin{equation}
	\phi_{h_i}(k_T,y)=\frac{C_F}{\alpha_s(2\pi)^3}\int d^2b d^2r e^{-i\bf{k}\cdot\bf{r}}\nabla^2_r\mathcal{N}_{h_i}^{g}(r,b,y)\,,
\end{equation}
where $\mathcal{N}_{h_i}^{g}(r,b,y)$ represents the dipole-hadron $h_i$ forward scattering amplitude for a gluon dipole with transverse size $r = r_T$ at impact parameter $b$. A uniform gluon distribution within the proton is assumed and the behavior of $\mathcal{N}_{h_i}^{g}$ is determined by solving the running coupling Balitsky-Kovchegov equation with the initial conditions from the AAMQS set~\cite{Albacete:2010sy}, exactly as done in~\cite{Albacete:2012xq}. 

The $k_T$-factorization formalism briefly presented above has been employed in phenomenological studies about hadron production over the years~\cite{Kharzeev:2001gp,Kharzeev:2002ei,Kharzeev:2004if,Levin:2010dw,Tribedy:2010ab,Tribedy:2011aa,Dumitru:2011wq,Albacete:2012xq,Duraes:2016yyg,Dumitru:2018gjm,Dumitru:2018yjs}. Folowing~\cite{Albacete:2012xq,Dumitru:2018gjm,Dumitru:2018yjs}, Equation \ref{Eq_fatkT} 
has been embedded into a Monte-Carlo Glauber simulation, as described in~\cite{ALbacete:2010ad,Albacete:2012xq}. Different from those studies, though, Equation \ref{Eq_fatkT} has been convoluted with the KKP Fragmentation Function model~\cite{Kniehl:2000fe}, $D_{g/h}$, in order to yield results at hadronic level even when considering momentum integrated obserbables and intrinsic saturation scale fluctuations~\cite{Iancu2005,Iancu:2004iy} were included when calculating particle multiplicities in a event-by-event basis. Qualitatively, this is inline with the expectation that, in the CGC framework, events with the largest multiplicities tend to be linked to rare color charge fluctuations in the colliding hadrons. Indeed, large event-by-event saturation scale fluctuations at nucleon level, driven by a log-normal distribution,
\begin{equation}
	P(\ln(Q^2_S/\langle Q^2_S\rangle))=\frac{1}{\sqrt{2\pi}\sigma}\exp{\left(-\frac{\ln^2(Q^2_S/\langle Q^2_S\rangle)}{2\sigma^2}\right)}\,,
\end{equation}
were found to be essential to describe multiplicity distributions within the CGC-based models~\cite{McLerran2016,Salazar:2021mpv,Terra2025,Ipp:2025sbt}. This expression generates an asymmetrical distribution of $Q_S/\langle Q_S\rangle$ around unity. The distribution of $Q_S$, 
as shown in Reference \cite{McLerran2016}, is significantly impacted by $\sigma$. In particular for multiplicity, $\sigma$ is crucial for the magnitude of multiplicity projections, especially for large $N$, i.e., for the tail, indicating how much we should increase or decrease the fluctuations of the non-perturbative saturation scale. 
Following study \cite{McLerran2016}, results with the $k_T$-factorization were obtained assuming $\sigma = 0.5$.

The results presented in this article were obtained from simulations using Pythia (Angantyr) version 8.313 and compared with the $k_T$-factorization. Collisions were simulated at different center-of-mass energies ($\sqrt{s}=$ 2.36, 5.02, 7.0, 13 TeV) and different pseudorapidity windows ($\eta<$ 0.5, 1.0, 2.0, 3.0). 

\section{Nucleus description: alpha-cluster and Woods-Saxon}
\label{sec3}

A complete understanding of the internal structure of nuclei is an open challenge in Nuclear Physics. A powerful tool for addressing this structure is ultrarelativistic heavy-ion collisions. 
In recent decades, experiments at the LHC and RHIC have brought successive advances in the characterization of heavy nuclei (especially lead), but many pieces of this puzzle are still missing. An important example is the spatial organization of lighter nuclei, such as oxygen. Understanding this requires investigating the influence of nuclei structure on observables generated in collisions between relativistic heavy ions.

The idea of the nucleus as composed of $\alpha$-particles appears in the literature not long after the discovery of the $\alpha$-particle itself. In 1920, E. Rutherford suggested that the nuclei of radioactive elements would be partly composed of helium nuclei ($\alpha$-particles) \cite{Rutherford1920}. However, the existence of a cluster structure in nuclei with 4n nucleons was only conceived a decade later in \cite{Gamow1930}, and subsequently the $\alpha$ clustering model for nuclei appears systematized and widely discussed in works such as, for example, Reference \cite{Bethe1936}, where the stability of the $\alpha$-particle is highlighted, suggesting that it can act as a building block in heavier nuclei. In contrast to the clustered view of the nucleus, the shell model proposes that the nucleus is composed of protons and neutrons, which are held together via the mean potential generated by nucleon-nucleon interactions. This model is highly effective in describing both the properties of the ground state and the excited states in nuclei, under the assumption that the building blocks are individual protons and neutrons. However, in the 1960s, excited states were identified in nuclei comprising equal numbers of protons and neutrons (e.g., $^{12}$C and $^{16}$O) that could not be (satisfactorily) described by the shell model. These states could then be associated with configurations composed of $\alpha$-particles. Molecular structures (clusters) appear systematically at the energy thresholds for decays. 
The fact is that studying collisions with oxygen makes it possible to probe the clustering model and investigate the 
possible effects of this cluster structure in smaller systems compared to larger nuclei.

\begin{figure}
\includegraphics[page=1,width=0.22\textwidth]{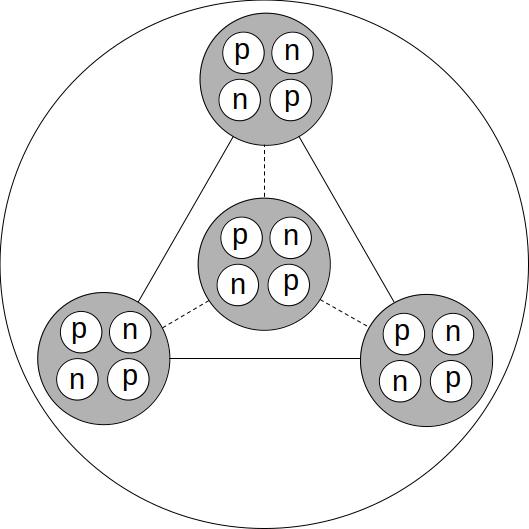}
\caption{Pictorial representation of the arrangement of $\alpha$-clusters within the oxygen nuclei.}
\label{fig_tetraedro}
\end{figure}

The oxygen is a light nucleus composed of 8 protons and 8 neutrons ($^{16}$O), 
being a natural candidate for a $\alpha$-clustered nucleus. When an $\alpha$-cluster structure is considered \cite{Wheeler1937,Hafstad1938,Huang2017}, the nucleons within the $^{16}$O would be arranged as four alpha particles forming a tetrahedral geometry. Such structure   
was obtained via Extended Quantum Molecular Dynamics (EQMD) -- based on the Quantum Molecular Dynamics (QMD) model -- with the effective Pauli potential \cite{Li2020}. According to the EQMD, the four clusters would be arranged at the vertices of a tetrahedron with an edge of 3.42 fm. 
Figure \ref{fig_tetraedro} shows the pictorial representation of the arrangement of nucleons within an $\alpha$ particle and the arrangement of $\alpha$-clusters within the nucleus of $^{16}$O. Studies involving ``clusterable'' nuclei, such as $^{16}$O, have the potential to explore the effects of the nuclear geometry of $\alpha$-clusters on the final state observables.

In Reference \cite{Frosch1967}, absolute cross sections for the elastic electron scattering of $^{4}$He were measured, and the authors identified evidence of a deviation from the Gaussian model, where the surface of the actual charge distribution appears less diffuse than that indicated by the Gaussian model with non-trivial central curvature. The authors indicate that in this case, the charge density can be well described by a three-parameter Fermi (3pF) distribution. More recently, the 3pF distribution has been applied, for example, in References \cite{Ding2023,Sharma2025,Behera2022}, to describe $\alpha$-clustering to evaluate the effects of the initial oxygen configurations on the observed final states. The nuclear charge density at a radial distance $r$ from the center of nucleus, via the 3pF distribution, can be written in the form
\begin{equation}
\rho(r)=\frac{\rho_0[1+w(r/R)]}{1+\exp{[(r-R)/a]}},
\label{Equation_3pF}
\end{equation}
where $\rho_0$ is the nuclear density at $r=0$, $R$ is the nuclear radius, $a$ is the diffuseness parameter and $w$ is the skin thickness parameter. For sampling the nucleon positions within the $\alpha$-clusters with Equation \ref{Equation_3pF}, we consider that 
each of the four $\alpha$-cluster has the values $R=0.964$ fm, $a=0.322$ fm, and $w=0.517$. Note that $w$ does not change the asymptotic behavior of the surface, and it would be enough to place four centers in a tetrahedron to create the nuclear cluster. Even with any reasonable internal density, the nucleus would already be clustered. However, $w$ is necessary because it allows for the non-Gaussian central curvature expected to occur in each cluster. It is the additional degree of freedom to make the central curvature more flexible. For each cluster, the positions of the nucleons are randomized by applying three successive Euler rotations, and to avoid overlap, a minimum distance of 0.9 fm is maintained between two nucleons within the same oxygen nucleus. In order to compare the effect of including clustering in oxygen, we also considered the Woods-Saxon distribution, which is a good approximation for describing nuclei whose shape could be that of a sphere, such as, for example, the lead nucleus. Therefore, it is a continuous distribution that provides an average density for the nucleus. This contrasts sharply with the idea of a clustered distribution. In practice, we take $w=0$ in the Equation \ref{Equation_3pF},
\begin{equation}
\rho(r)=\frac{\rho_0}{1+\exp{[(r-R)/a]}}.
\label{Equation_2pF}
\end{equation}
The parameter values used in Woods-Saxon are $R=2.608$ fm, $a=0.513$ fm and $w=0$.

\begin{figure}
\includegraphics[page=1,width=0.49\textwidth]{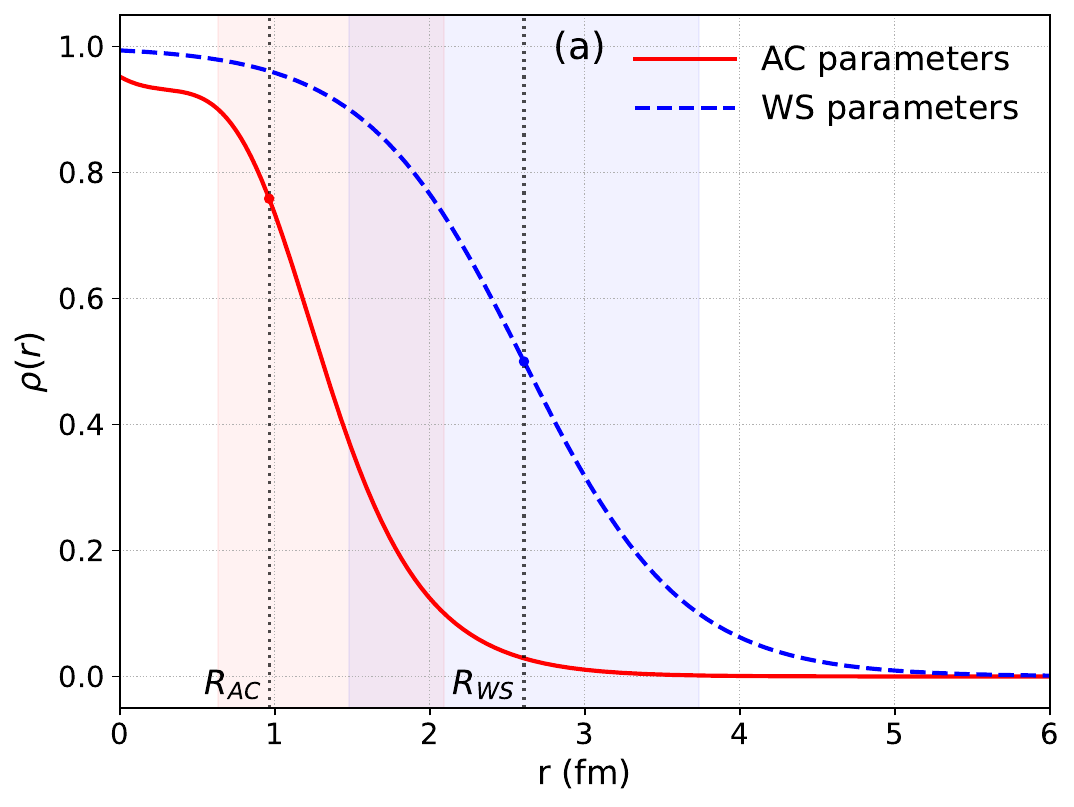}
\includegraphics[page=1,width=0.50\textwidth]{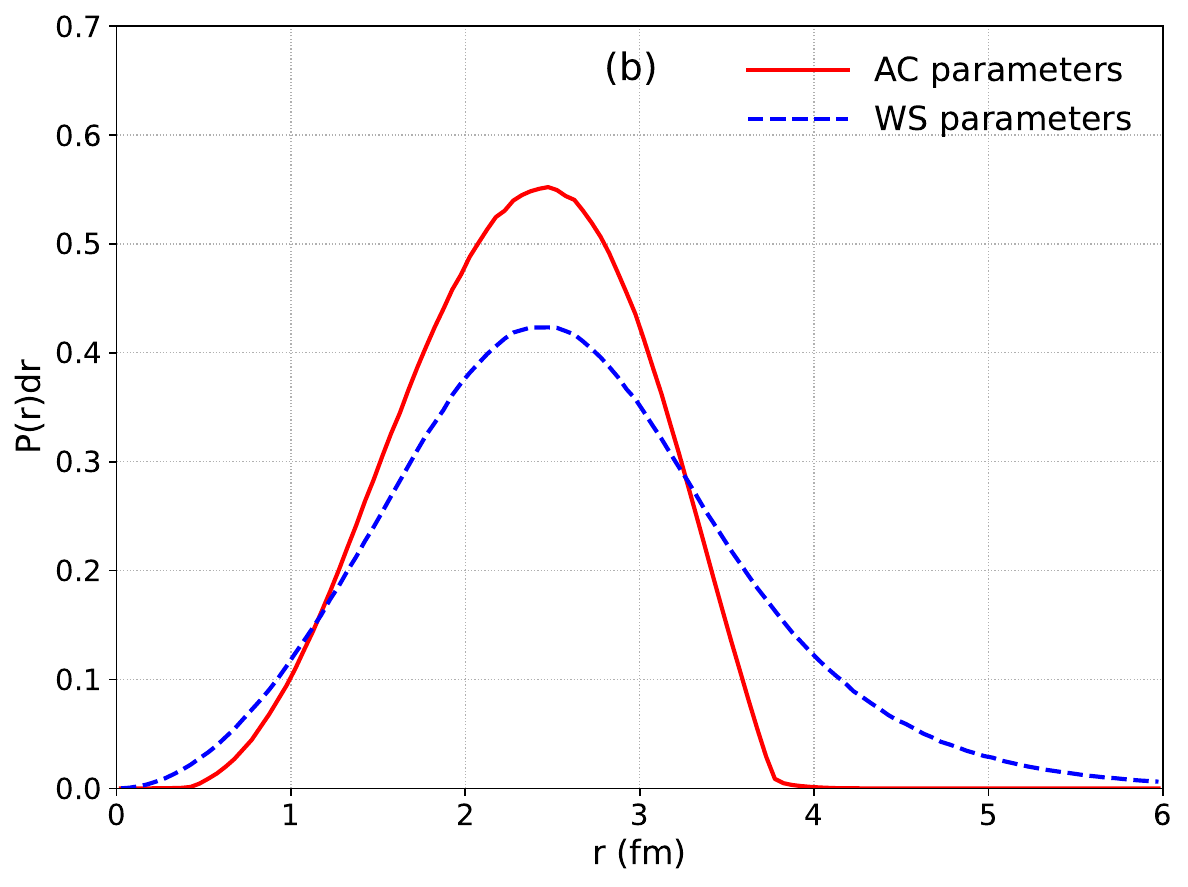}
\caption{Comparison between the parameter sets for the $\alpha$-cluster and Woods-Saxon models: (a) density profile for each parameter set; (b) probability of the radial position of nucleons within the oxygen nucleus.}
\label{fig_nucleodescricao}
\end{figure}

In morphological terms, as Figure \ref{fig_nucleodescricao}(a) shows -- and as expected -- the nucleus described by the Woods-Saxon parameter group (here named ``WS parameters'') is more extensive than a single cluster described by the $\alpha$-cluster parameter group (here named `AC parameters''). The fact that Woods-Saxon nuclear radius ($R_{WS}$) crosses the center of its diffusion region -- radial band where the nuclear density gradually decreases from the interior value to zero on the exterior, controlled by parameter $a$ -- (blue shaded area) and $\alpha$-cluster nuclear radius ($R_{AC}$) shifts to the left of the center of its diffusion region (red shaded area) reveals a fundamental difference between these models: the parameter $R$ only coincides with the middle of the diffusion region if we take $w=0$, on the other hand, when we take $w\neq 0$ the term $w(r/R)^2$ survives and shifts the point where the density drops to 50\%, therefore, in this case $R$ crosses the diffusion region not at the center, showing less density at the center and more radial weight at the surface. Figure \ref{fig_nucleodescricao}(b) shows the comparison of the probability of the radial position of nucleons within the oxygen nucleus for the two different sets of parameters. 
The radial distribution of individual nucleons was calculated event by event after the complete geometric construction of the nucleus, for both sets of parameters.
As expected, AC parameters presents a more compact radial distribution of nucleons compared to WS.

\section{Charged-particle multiplicity distribution}
\label{sec4}

\begin{figure}
\includegraphics[page=1,width=0.49\textwidth]{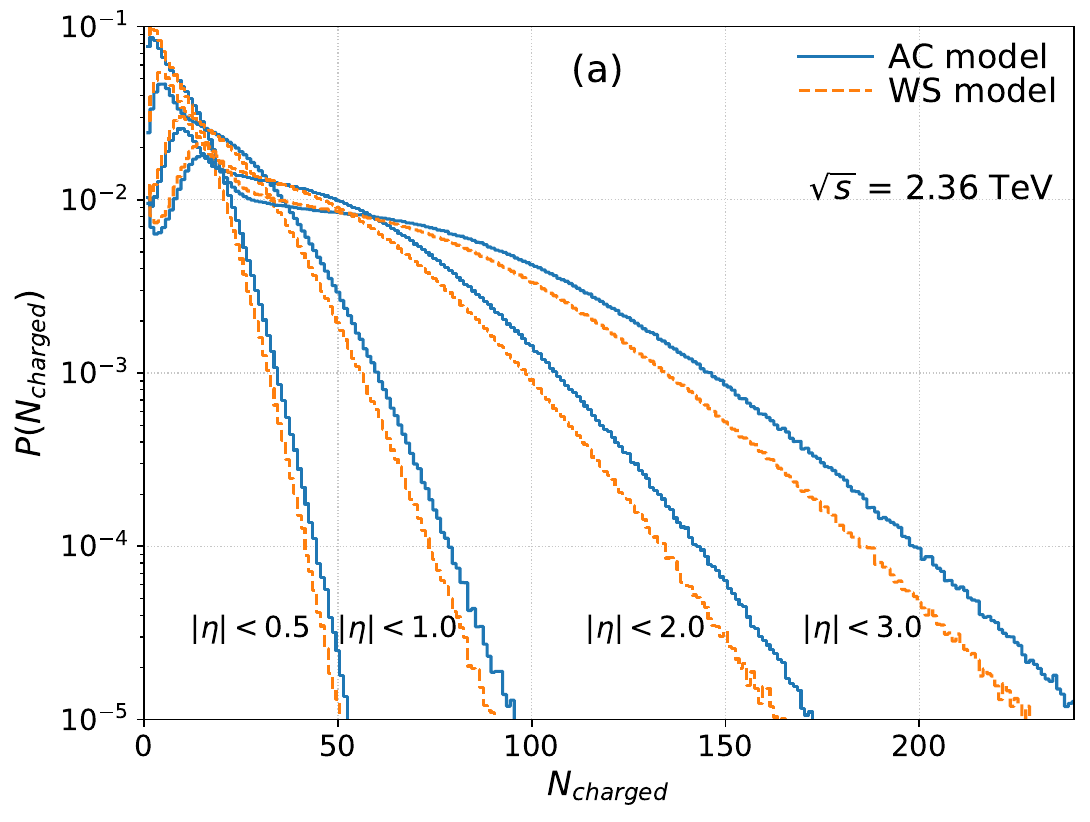}
\includegraphics[page=2,width=0.49\textwidth]{plotMultiplicityGeometryComparison.pdf}
\includegraphics[page=3,width=0.49\textwidth]{plotMultiplicityGeometryComparison.pdf}
\includegraphics[page=4,width=0.49\textwidth]{plotMultiplicityGeometryComparison.pdf}
\caption{Charged particle multiplicity in p-O collisions. Each panel shows a different center-of-mass energy ($\sqrt{s} = 2.36, 5.02, 9.62, 13$ TeV) and compares the Alpha Cluster (AC) and Woods-Saxon (WS) models at different pseudorapidity values ($|\eta| < 0.5, 1.0, 2.0, 3.0$).}
\label{fig_multiplicitypO}
\end{figure}

Figure \ref{fig_multiplicitypO} shows the multiplicity of charged particles produced in $p$-O collisions considering the previously discussed nuclear description models: the $\alpha$-cluster model (here named ``AC model'') and the Woods-Saxon model (here named ``WS model''). Each panel shows a different center-of-mass energy ($\sqrt{s} = 2.36, 5.02, 9.62, 13$ TeV) and compares the two models in different pseudorapidity windows ($|\eta| < 0.5, 1.0, 2.0, 3.0$). These distributions were obtained using Pythia (Angantyr). 
In all cases, we can see a clear difference between the models across the entire observed spectrum; this difference gradually increases as we consider higher values of $N_{charged}$. The AC model curve starts slightly below the WS model, but an inversion soon occurs. This inversion shifts towards higher $N_{charged}$ as we consider higher pseudorapidity. The small difference observed at low multiplicity indicates a low sensitivity of this region to the geometry of the considered nucleus; here, the ``average geometry'' dominates because the proton is expected to interact with few nucleons, and the models roughly coincide within the statistical uncertainty.  
On the other hand, the difference between the AC and WS models becomes increasingly pronounced as $N_{charged}$ increases and also as we consider a larger range of pseudorapidity. 
This difference is a consequence of how the organization of the nuclear substructure enters the fluctuations of the initial state and propagates to the multiplicity observed at the end. The tails of the distributions reflect, mainly, the rarer events, where there is a high number of participating nucleons, or the initial energy density was very high. In this case, the tail is more sensitive to extreme fluctuations in the initial geometry. In the AC model, nucleons are grouped into coherent objects (clusters), and each cluster has a high local density in a small region of space. This is radically different from the WS model, where nucleons are distributed in an approximately continuous and poorly correlated manner. This results in the difference observed most markedly in the tail of the panels in Figure \ref{fig_multiplicitypO}. Therefore, we see here a way to differentiate nuclear geometry models for oxygen using a global inclusive observable.

\begin{figure}
\includegraphics[page=1,width=0.49\textwidth]{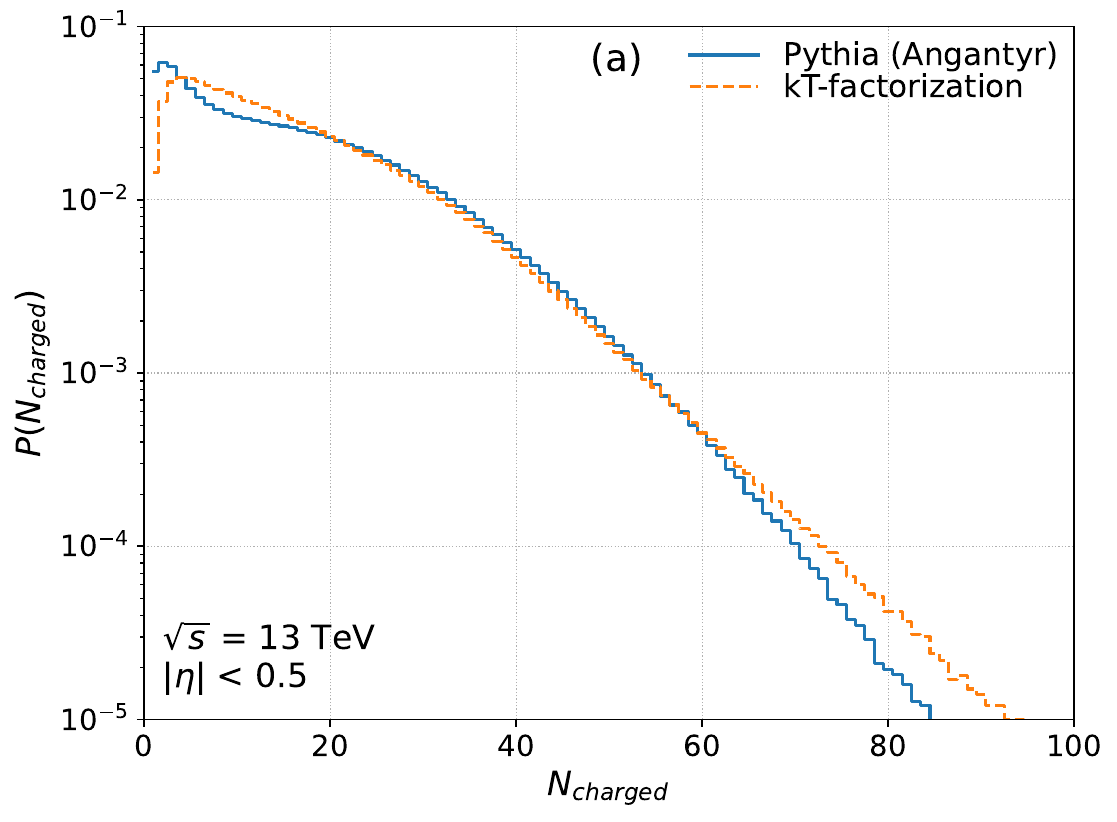}
\includegraphics[page=4,width=0.49\textwidth]{plotFactorizationsComparison.pdf}
\caption{Charged particles multiplicity in p-O collisions. Comparison between the Pythia (Angantyr) and the $k_T$-factorization considering $\sqrt{s} = 13$ TeV and (a) $|\eta| < 0.5$ and (b) $|\eta| < 3.0$.}
\label{fig_factorizationscomparison}
\end{figure}

In Figure \ref{fig_factorizationscomparison} we compare the charged particles multiplicity calculated with Pythia and the $k_T$-factorization for a center-of-mass energy of 13 TeV, considering the $\alpha$-cluster distribution for the $^{16}$O nucleus. In panel \ref{fig_factorizationscomparison}(a) this calculations are performed for $|\eta| < 0.5$ and in panel \ref{fig_factorizationscomparison}(b) for $|\eta| < 3.0$. We can see that there are significant differences between the results obtained with with Pythia and with the $k_T$-factorization. At small $N_{charged}$, the $k_T$-factorization does not present the same structure (peak-valley) that appears in the Pythia simulations. At intermediate $N_{charged}$ the curves approximately coincide at $|\eta|<0.5$, but diverge at $|\eta|<3.0$. In the tail, there is significant divergence in both panels. In the future, we will be able to compare these formalisms with experimental data and then understand what kind of physical processes best describe the data for these configurations. 

\begin{figure}
\includegraphics[page=1,width=0.49\textwidth]{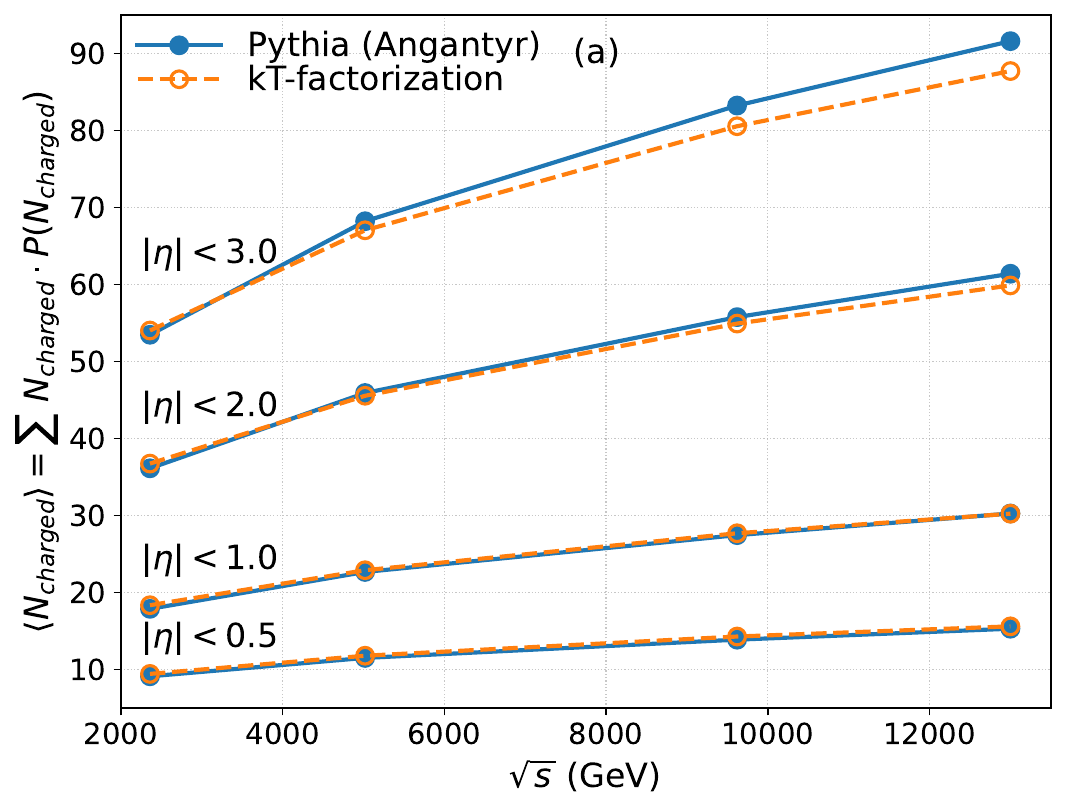}
\includegraphics[page=1,width=0.49\textwidth]{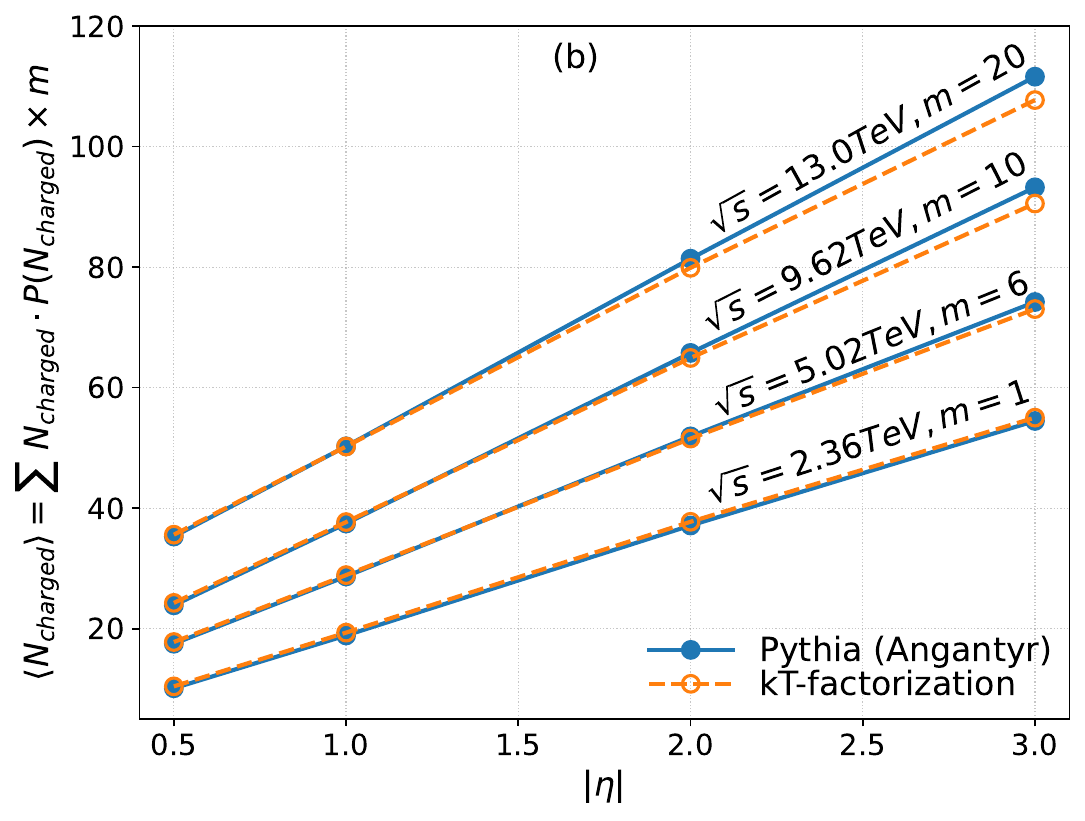}
\caption{Average charged particle multiplicity (a) as a function of the center-of-mass energy and (b) as a function of pseudorapidity. Comparison between the Pythia (Angantyr) and the $k_T$-factorization considering different center-of-mass energy ($\sqrt{s} = 2.36, 5.02, 9.62, 13$ TeV) and pseudorapidity values ($|\eta| < 0.5, 1.0, 2.0, 3.0$). Figure (b) we multiply by the factor $m = 1.0, 6.0, 10.0, 20.0$ to make it easier to visualize.}
\label{fig_Naverage}
\end{figure}

Using the charged particles multiplicity calculated with Pythia (Angantyr) and $k_T$-factorization, we can calculate the average multiplicity $\langle N_{charged} \rangle$. It is interesting to observe the behavior of $\langle N_{charged} \rangle$ in terms of its evolution with energy, because this allows us to verify if the model used reproduces the expected energy growth -- a non-abrupt growth with increasing energy -- in the different pseudorapidity windows. Furthermore, this presents the possibility of separating longitudinal effects from geometric effects.  Figure \ref{fig_Naverage}(a) shows $\langle N_{charged} \rangle$ as a function of the energy $\sqrt{s}$. We can see that both approaches provide fairly similar averages for more central pseudorapidity and differ (even slightly) in more forward pseudorapidity and higher energy. We can also note that the average increases with energy and with the value of pseudorapidity considered. Figure \ref{fig_Naverage}(b) rearranges the same information now as a function of $\eta$ with the center-of-mass energy fixed. We multiply by the factor $m = 1.0, 6.0, 10.0, 20.0$ to make it easier to visualize. Note that in this way we have an indirect probe of how particle production is distributed longitudinally while keeping the center-of-mass energy constant; therefore, in this case, the integrated production responds to the transition in $x$. Observe that here, as we are dealing with multiple scales of $p_T$ and different production mechanisms, each point $\eta$ does not correspond to a single value of $x$ but rather to a range of $x$. In hadronic processes, the typical kinematic relationship is $x_{a,b}=(p_T/\sqrt{s})e^{\pm y}$, so the reorganization given in Figure \ref{fig_Naverage}(b) reflects the cumulative contribution of progressively smaller and more asymmetric Bjorken-x regions, where varying the value of the pseudorapidity $\eta$ causes the value of $x$ to changes strongly.

\section{KNO scaling in charged-particle multiplicity distributions}
\label{sec5}

A key feature of multiplicity distributions is the Koba-Nielsen-Olesen (KNO) scaling \cite{KNS1972}, a concept originally derived from Feynman scaling. It suggests that at asymptotically high energies, multiplicity distributions exhibit a universal scaling behavior. Specifically, the particle production probability in a particular phase space can be expressed as
\begin{equation}
P(N(s)) = \frac{1}{\langle N(s) \rangle} \, \Psi\left( \frac{N}{\langle N(s) \rangle} \right) + \mathcal{O}\left(\frac{1}{\langle N(s) \rangle^2}\right).
\end{equation}
The first term is the dominant term and carries the idea of self-similarity in particle production -- in practice, this is the part that interests us in this work -- and the second are the higher-order subdominant corrections. Furthermore, \( \langle N(s) \rangle \) is the average multiplicity at center-of-mass energy \( \sqrt{s} \) and $\Psi(z)$, with \( z = N_{charged}/\langle N_{charged} \rangle \), is an energy-independent (universal) function. Therefore, under KNO scaling hypothesis, the shape of the scaled distribution becomes independent of energy. The multiplicity distributions at different energies collapse onto a single universal curve when plotted as a function of the scaled variable $z$. For a comprehensive discussion of KNO scaling, see Reference \cite{Grosse2010}.

KNO scaling in hadronic collisions is commonly and continuously addressed in theoretical, phenomenological, and experimental investigations. Recent research highlights both the robustness and limitations of this principle of universality, outlining the context for the results presented here. 
From a theoretical point of view, it is important to try to relate the different properties of the KNO scaling with the dynamics of quarks and gluons in the high-density 
regime. 
For example, in the References \cite{DumitruNara2012,DumitruPetreska2012}, the authors show that KNO scaling results, on a scale of the order of the saturation moment $Q_s$, when the theory describing color charge fluctuations is approximately Gaussian, and both saturation as well as running coupling effects are necessary for KNO scaling to occur. In Reference \cite{Terra2025}, using the \(k_T\)-factorization structure, the authors seek to access the initial geometric shape of the proton (spatial shape at the moment of collision) in $p$-Pb collisions and calculate the multiplicity distributions. They observed that 
combining a baryon junction geometry for the nucleon with 
dynamic fluctuations of the saturation scale provided an approximate description of the multiplicity distributions in $p$-Pb collisions. Another interesting discussion is found in Reference \cite{MartinsFontes2025PRD,MartinsFontes2025Physics} where, considering an explicit separation between soft and semi-hard processes in the $k_T$-factorization approach, the authors observed that, in $pp$ collisions and within the central pseudorapidity region, the KNO scaling is valid. From a phenomenological perspective, the authors of Reference \cite{Mahmoud2022} systematically tested the KNO scale in $pp$ collisions using Pythia simulations at different center-of-mass energies and pseudorapidity ranges. The KNO scale was reported to be approximately valid for $\eta<$ 0.5, but significant violations appear as we consider larger pseudorapidity values, especially for large $N_{charged}$, implying increasing fluctuations and a breakdown of universality in the particle production mechanism. References, for example, \cite{ALICE2017, ALICE2025}, on the other hand, present experimental data on the multiplicity of charged particles in a wide pseudorapidity window for pp and $p$-Pb collisions and elaborate on the KNO scaling of these data. 

\begin{figure}
\includegraphics[page=1,width=0.49\textwidth]{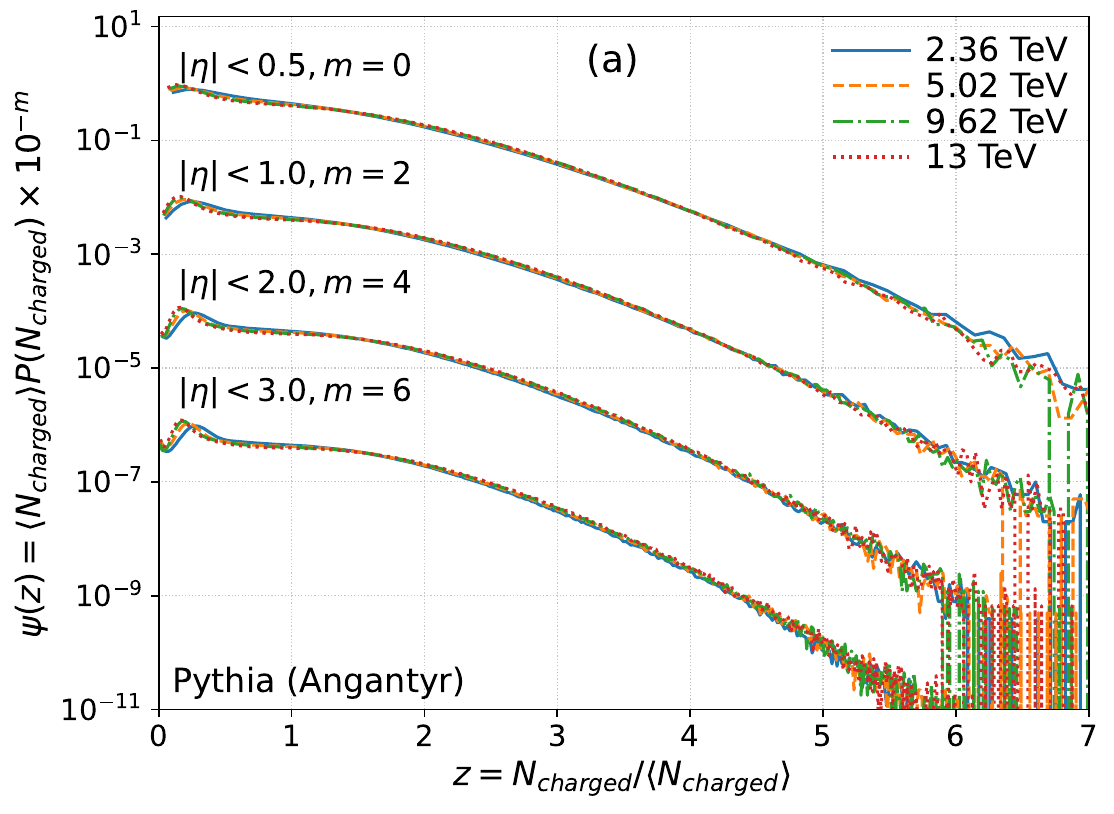}
\includegraphics[page=1,width=0.49\textwidth]{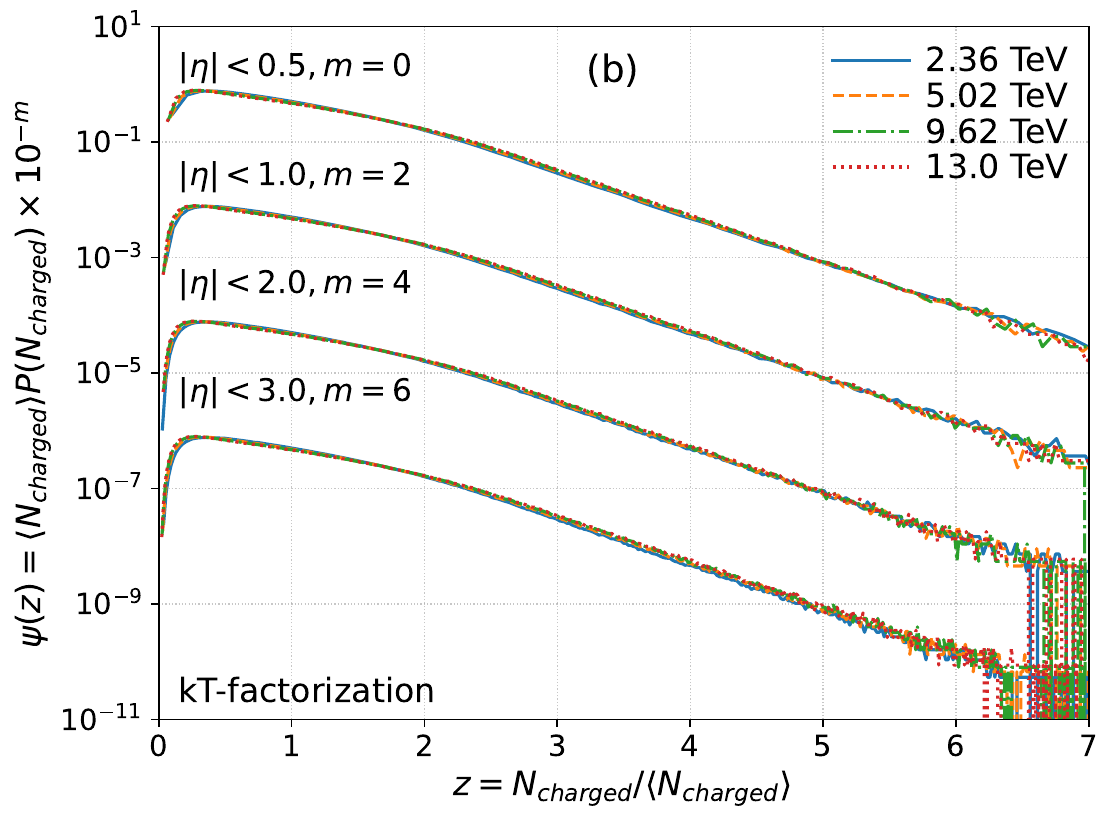}
\caption{Charged particle multiplicity distribution on the KNO scale for $p$-O collisions. Comparison of different center-of-mass energies ($\sqrt{s}=$ 2.36, 5.02, 9.62, 13.0 TeV) and pseudorapidities ($\eta<$ 0.5, 1.0, 2.0, 3.0) in (a) collinear factorization and (b) $k_T$-factorization. The sets of curves (per $\eta$) were multiplied by a factor of $10^{-m}$ to facilitate visualization.}
\label{fig_KNOscaling}
\end{figure}

In Figure \ref{fig_KNOscaling} the multiplicity distributions obtained were scaled following the KNO form for different center-of-mass energy configurations ($\sqrt{s}=$ 2.36, 5.02, 9.62, 13.0 TeV) and pseudorapidities ($\eta<$ 0.5, 1.0, 2.0, 3.0), the sets of curves (by $\eta$) were multiplied by a factor of $10^{-m}$ to facilitate visualization. Panel (a) shows the results for the Pythia (Angantyr) simulations and panel (b) for the $k_T$-factorization. We can see that in both panels, for all configurations, there are no significant divergences between the curves. Multiplicity distributions at different energies approximately collapse onto the same curve without new scaling as the energy increases.

\section{Parameterization with double NBDs}
\label{sec6}

Determining which mathematical function can describe a multiplicity distribution can contribute to understanding the underlying physical mechanism of particle production. Regarding the multiplicity of charged particles -- up to a certain limit -- the Negative Binomial Distribution (NBD) proves to be a good fit \cite{UA51988gup,UA51985PLB167}. The NBD emerges directly from an intuitive physical model and is able to ``encapsulate'' the physics of particle production at high energies. The fact that the NBD is successful in describing the data is a strong indication that particle production is not an independent process, pointing to the presence of correlations between the final particles. This discussion is addressed in detail, for example, in the References \cite{UA51985PLB167,UA51985PLB160,Grosse2010}. NBD is defined as
\begin{equation}
P_{NBD}(n,p,k)=
\begin{pmatrix}
n+k-1\\
n
\end{pmatrix}
(1-p)^n\ p^k.
\label{NBDsimples}
\end{equation}
where $k$ measures the degree of clustering: for $k\to\infty$, the NBD tends towards a Poisson distribution (processes with independent events); a high value of $k$ indicates that the correlations between the particles produced are very weak and the distribution tends towards a Poisson; for $k=1$, the NBD becomes a Geometric distribution. 
Equation \ref{NBDsimples} answers the question of what is the probability of $N_{charged}$ failures occurring before the $k$-th success occurs. 
Considering negative integer values of $k$ and $p^{-1}=1+<n>/k$, the generalization results in
\begin{equation}
P_{NBD}(n,\langle{n}\rangle,k) = \frac{(n+k-1)!}{n!(k-1)!} (1-\langle{n}\rangle)^k \langle{n}\rangle^n.
\label{NBDfunc}
\end{equation}

The descriptive power of the NBD becomes limited at high energies. For example, in Reference \cite{Giovannini1999}, deviations between the multiplicity distribution and the fit with a single NBD are found in pp collisions at $\sqrt{s}$ = 900 GeV. In this case, a combination of two NBDs \cite{Giovannini1999} has proven efficient in reproducing current experimental data. Each component of the double NBD corresponds to a distinct class of events, commonly called ``soft'' and ``semi-hard'' processes. This is discussed, for example, in \cite{Giovannini1999,Ghosh2012,Zborovsky2018} and, more recently, in \cite{MartinsFontes2025PRD,MartinsFontes2025Physics}. This classification is based on different types of events, and not on distinct mechanisms of particle production within a single event. As a result, the overall multiplicity distribution can be written as
\begin{equation}
P(n) = \alpha_{soft} P_{NBD}(n,\langle{n}\rangle,k_1)  + (1-\alpha_{soft})P_{NBD}(m,\langle{n}\rangle,k_2) \nonumber 
\label{eq_DNBD}
\end{equation}
The double NBD is then fitted to the multiplicity distributions measured in this analysis. Since $N\sim 0$ has considerable background noise from diffractive interactions, we take the distributions with for $N\geq 1$.

\begin{figure}
\includegraphics[page=1,width=0.49\textwidth]{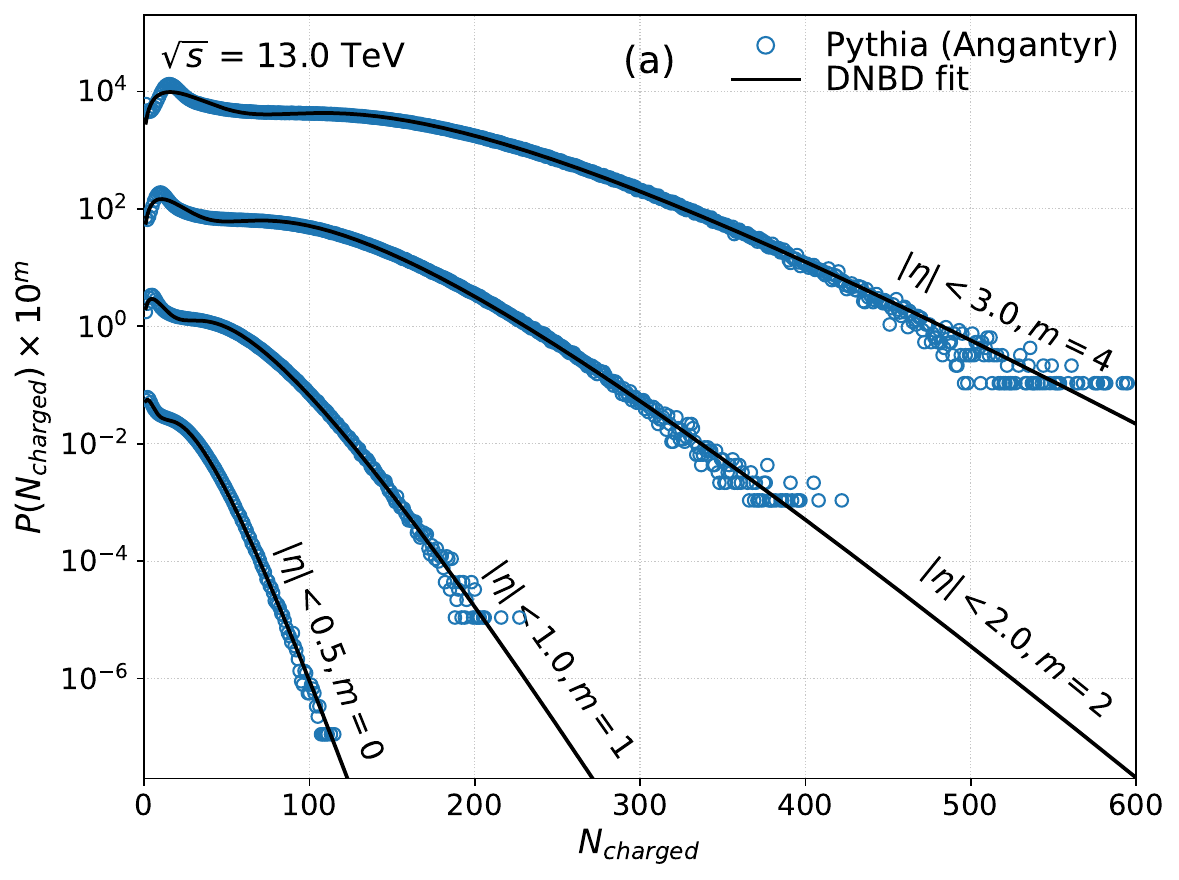}
\includegraphics[page=1,width=0.49\textwidth]{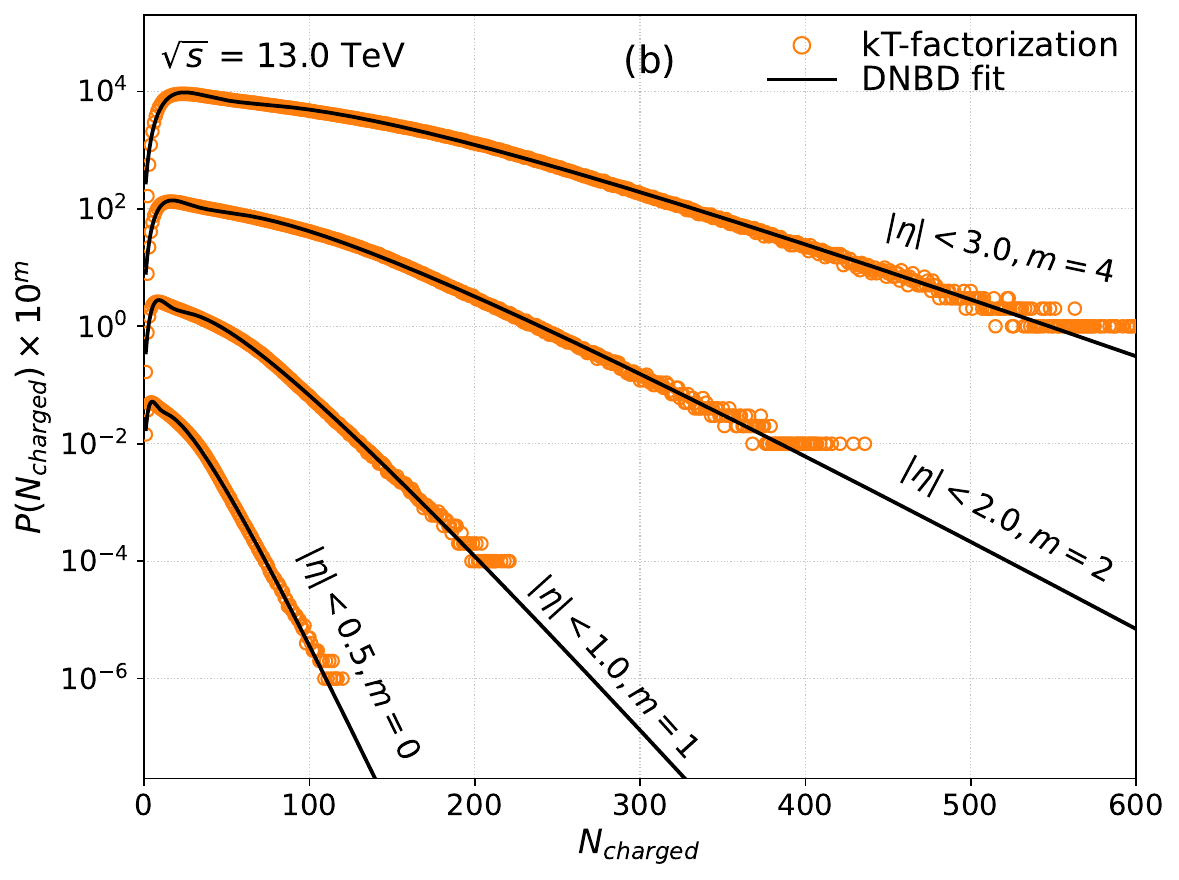}
\caption{Charged-particle multiplicity distribution. Comparison of the Pythia (panel (a)) and $k_T$-factorization (panel (b)) models with the DNBD fit for $\sqrt{s}=$ 13.0 TeV observing different pseudorapidity ranges ($\eta<$ 0.5, 1.0, 2.0, 3.0) in (a).}
\label{fig_2NBD}
\end{figure}

In Figure \ref{fig_2NBD} we compare the charged particle multiplicity distribution, discussed in previous sections, with the fit performed with two NBDs, as defined in Equation \ref{eq_DNBD}. We consider different pseudorapidity values ($|\eta| <$ 0.5, 1.0, 2.0, 3.0) and center-of-mass energy values ($\sqrt{s} =$ 2.36, 5.02, 7.0, 13.0 TeV). It can be observed that the double NBD fits the multiplicity distributions well in all cases analyzed. In particular, for small values of $N_{charged}$, an approximately Poissonian distribution originating from the first NBD is observed, and the long tail, which reaches the part of the distribution with lower statistics, comes from the second NBD. The fact that the NBD pair fits the multiplicity distributions points to, as expected, the presence of different classes of events (soft and semi-hard) that overlap and result in the observed multiplicity distribution, each of these classes having its own particular production dynamics and statistical characteristics. Each NBD captures one of these classes of events, and their sum indicates that the production of charged particles in $p$-O collisions is not statistically homogeneous.

\section{Conclusion}

In this work, we used the Pythia (Angantyr) event generator and $k_T$-factorization to study the production of charged particles in $p$-O collisions at the LHC. We considered the structure of the oxygen nucleus through the $\alpha$-cluster and Woods-Saxon models. Our focus was on the multiplicity distribution of charged particles and our main conclusions are:

\vskip0.2cm
\noindent
- The multiplicity of charged particles is strongly influenced by the initial geometry considered. The geometric description of the oxygen nucleus as being $\alpha$-clusters arranged at the vertices of a tetrahedron or a Woods-Saxon projects significantly different multiplicities, particularly for large $N_{charged}$ and more expressively at higher pseudorapidity.

\vskip0.2cm
\noindent
- The multiplicity of charged particles calculated with Pythia (Angantyr) reveals significantly different behavior from that calculated with $k_T$-factorization. The characteristic peak-valley structure of Pythia at small $N_{charged}$ does not appear in the $k_T$-factorization calculations. Large $N_{charged}$ also shows a large divergence between the models.

\vskip0.2cm
\noindent
- We observed that in all cases there is universality in the behavior of the multiplicity, that is, the KNO scale is not violated. 

\vskip0.2cm
\noindent
- We also observed that the DNBD fit describes the projections well with Pythia (Angantyr) and also with $k_T$-factorization.

Based on the current data obtained from $p$-O collisions at the LHC, these findings contribute to an enhanced understanding of charged particle multiplicity distribution measurements and offer potential insights into mechanisms of particle production under these conditions, contributing to a more comprehensive understanding of the dynamics governing high-energy nuclear interactions and the role of system size in shaping final-state observables.


\section{ACKNOWLEDGEMENTS}

This study was supported, in part, by FAPESP (contract numbers 2024/17836-9, 2024/14337-1, and 2020/04867-2), and by INCT CERN-Brasil (grant number 406672/2022-9). AVG acknowledges support from CNPq through the INCT-FNA grant 408419/2024-5 and the Universal Grant 405458/2025-8 and the National Laboratory for Scientific Computing (LNCC/MCTI, Brazil), through the ambassador program (UFGD, subproject SADFT), for providing HPC resources of the SDumont supercomputer.


\end{document}